# THREAD REVIEW SENTIMENTAL ANALYSIS WITH TKINTER GUI & TABLEAU DASHBOARD

[1]Robin Donal, [2] Prof. Soma Joshi

[1]Student, [2] Assistant Professor

[1]Department of Information Technology,

[1]Ajeenkya DY Patil University, Pune, India

Abstract: This project focuses on utilizing a combination of Tkinter for GUI development and Tableauf for data visualization to do sentiment analysis on thread reviews.The main goal is to evaluate and visualize consumer sentiments as they are expressed in thread reviews in order to provide insights into areas for improvement, preferences, and customer satisfaction.The procedure starts with gathering thread reviews from many sources, which are then cleaned and prepared for analysis through preprocessing.Sentiment analysis classifies opinions as good, negative, or neutral based on the expressed sentiment by applying natural language processing techniques.The standard Python GUI package Tkinter is used to create an interactive user interface that allows users to enter thread reviews, start the sentiment analysis process, and see the analysis's outcomes.With the help of the user-friendly GUI, users may interact with the system and acquire insightful information with ease.Additionally, Tableau is used to produce a dynamic and eye-catching dashboard that displays the findings of the sentiment analysis using a variety of charts and graphs.Stakeholders may make educated decisions based on the studied data by using the dashboard, which provides a thorough overview of the sentiment distribution, frequency of positive and negative reviews, trending topics, and other pertinent indicators.Overall, this project offers a solid method for analyzing and comprehending customers' sentiments from thread reviews by integrating Tableauf for GUI development with Tkinter for sentiment analysis and data visualization. This allows for the creation of meaningful dashboards.

*Index Terms* - Thread, Sentiment, Emotion, Market.

## 1.INTRODUCTION

Thread review sentiment analysis using Tableau Dashboard and Tkinter GUI is a thorough method for interpreting and analyzing user reviews from a variety of sites.This project combines the power of natural language processing (NLP) for sentiment analysis, a dynamic visualization tool with Tableauf for intelligent data representation, and a user-friendly graphical interface using Tkinter for simplicity of use.The sentiment analysis component analyzes textual data from reviews to ascertain the users' expressed sentiment (positive, negative, or neutral).TkinterGUI offers an interactive interface that enables users to easily enter data, begin analyses, and display the outcomes.However, the Tableau Dashboard provides interactive and aesthetically pleasing representations that help stakeholders understand the significance of the sentiment analysis data.Businesses and people are empowered by this comprehensive solution to better understand customer attitudes, make educated decisions, and improve user experiences. Data collection: Compiling user reviews from a range of websites or platforms whereCustomers provide comments on goods or services. Text Preprocessing: removing extraneous letters, numerals, and stopwords from the acquired text data and tokenizing it into meaningful words. Sentiment analysis: classifying each review's sentiment as good, negative, or neutral using NLP algorithms.

Deep learning models like Recurrent Neural Networks (RNNs) and Transformers, or machine learning methods like Naive Bayes and Support Vector Machines, can be used to accomplish this classification.

Tkinter GUI: Using Tkinter, a user-friendly interface is created that enables users to submit new reviews, start sentiment analysis, and examine the results in real time.Using Tableau, create an interactive dashboard to visualize the sentiment analysis results.Key indicators, including the distribution of good, negative, and neutral reviews over time, the top positive and negative phrases, and general sentiment patterns, can be seen on the dashboard.

## 2. LITERATURE REVIEW.

The body of research on sentiment analysis in thread reviews using TkinterGUI and Tableau Dashboard indicates that there is growing interest in using data visualization and natural language processing (NLP) approaches to assess and comprehend the sentiment expressed in online discussions.Academics likeIn order to accurately classify text sentiment, Liu (2012), Pan and Lee (2008), and others explored a variety of machine learning algorithms and lexicon-based approaches. Their work paved the way for more advanced sentiment analysis methods, such as deep learning models like convolutional neural networks (CNNs) and recurrent neural networks (RNNs), which have demonstrated superior performance in sentiment classification tasks (Kim, 2014; Tang et al., 2015).Furthermore, scientists have emphasized the significance of preprocessing methods such tokenization, stemming, and stopword removal in enhancing the accuracy of sentiment analysis (Bird, 2006; Manning et al., 2008).In the meantime, research aimed at improving the usability and accessibility of sentiment analysis tools has highlighted the incorporation of user-friendly interfaces, such as TkinterGUIs, for data entry and interaction (Dixetal.,2004;Burnettetal., 2014).Additionally, researchers have acknowledged the value of data visualization in efficiently communicating sentiment analysis results. For example, TableauDashboard allows users to graphically investigate sentiment trends, patterns, and outliers (Heer and Agrawala, 2006; Segel and Heer, 2010).The literature emphasizes the multidisciplinary character of sentiment analysis in thread reviews, utilizing knowledge from data visualization, computer science, linguistics, and human-computer interaction to create comprehensive methods for comprehending and interpreting sentiment within online threads.Even while previous research has significantly improved sentiment analysis algorithms and interface design, more research is still needed to solve issues like sarcasm.context-aware sentiment analysis, real-time sentiment dynamics visualization with in-progress conversations, and mood recognition.

## 3. SYSTEM ARCHITECTURE

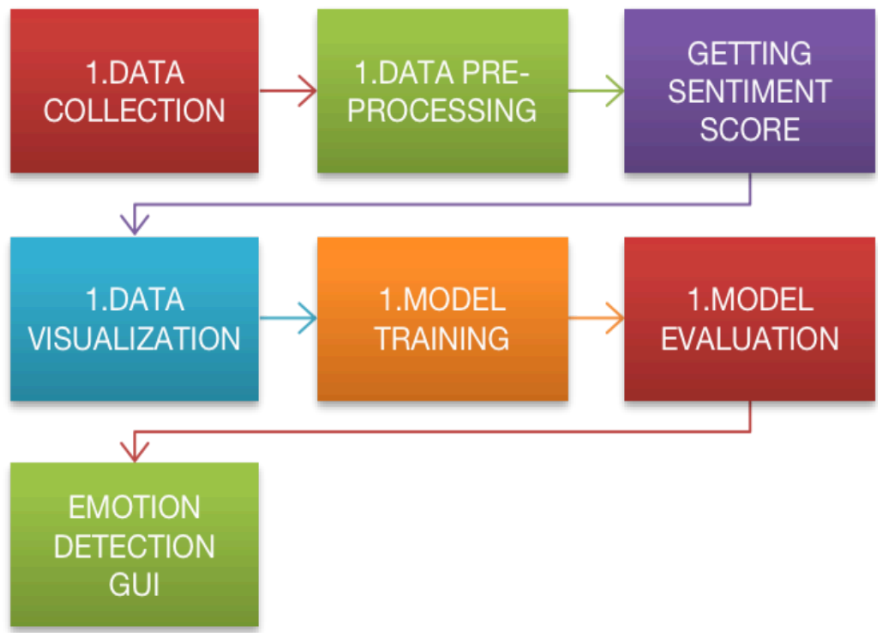

Fig 1: System architecture

## 4. METHODOLOGY

4.1. Data Collection

a. DataSources: Determine which internet forums, social media platforms, and product review websites are good places to start gathering thread reviews.

b. Data Extraction:- To extract thread reviews data, use web scraping methods or API calls. Save the extracted data in an organized format (such as Excel or CSV).

4.2. Data Preprocessing

a. Data Cleaning:- Took out duplicates, unnecessary information, and noise from the collection. Appropriately handle missing values (e.g., imputation, removal).

b. Text Preprocessing:- Tokenization (splitting reviews into word-sort tokens). Lowercasing: To ensure consistency, all text is converted to lowercase. Eliminating common words that lack significant meaning is known as stopword removal. Lemmatization/stemming: reducing words to their most basic form.

4.3. Sentiment Analysis

a. Sentiment Classification:- Classify reviews into positive, negative, or neutral sentiments using pre-trained sentiment analysis models or custom-built machine learning algorithms.

b. Sentiment Scoring:- Using a scale of 1 to 5 or -1 to 1, assign sentiment corestore views.
Development of GUID using Tkinter

4.4. GUI Development with Tkinter

a. Interface Design:- Design a user-friendly interface using Tkinter to allow users to input text for sentiment analysis.

b. Integration:- Integrate the sentiment analysis functionality into the Tkinter GUI.
Display the sentiment classification results to the user.

4.5. Data Visualization with Tableau

a. Data Preparation:- Export data to a CSV file or prepare the cleaned and processed data for Tableau visualization.

b. Dashboard Design:- Make dynamic dashboards in Table that automatically visualize sentiment analysis findings. Incorporate visual aids like word clouds, pie charts, and bar charts to display sentiment distributions, most-frequent words, and other data.

c. User Interaction:- Make sure the Tableau dashboard is easy to use and permits interactive data exploration (such as filtering and sorting).

4.6. Analysis and Interpretation

a. Insights Extraction:- Examine the sentiment analysis findings to derive significant insights regarding the thread reviews.Determine the main themes, patterns, and trends in the views.

b. Reporting:- Using Tableau visualizations and sentiment analysis insights, compile the results into an extensive report.Present the port in an easy-to-understand manner.

4.7. Validation and Testing

a. Accuracy Assessment:- Determine the sentiment analysis model's accuracy by contrasting its predictions with manually labeled data, if accessible.

b. Usability Testing:- Perform usability testing on the Tableau Dashboard and Tkinter GUI to find any problems or areas that need work.

4.8. Deployment

a. Deployment of GUI:- Give the intended users or stakeholders access to the TkinterGUI application.

b. Deployment of Tableau Dashboard:- Host the Tableau Dashboard on a server or platform that is appropriate for user access.This methodology offers an organized way to do sentiment analysis on thread reviews, illustrate the outcomes with a Tableau dashboard, and effectively explain the findings.Modifications

## 5. RESULTS AND DISCUSSION
### 5.1 MODEL ACCURACY AND CONFUSION MATRIX

```
              precision    recall  f1-score   support

           1       1.00      1.00      1.00         1

    accuracy                           1.00         1
   macro avg       1.00      1.00      1.00         1
weighted avg       1.00      1.00      1.00         1
```

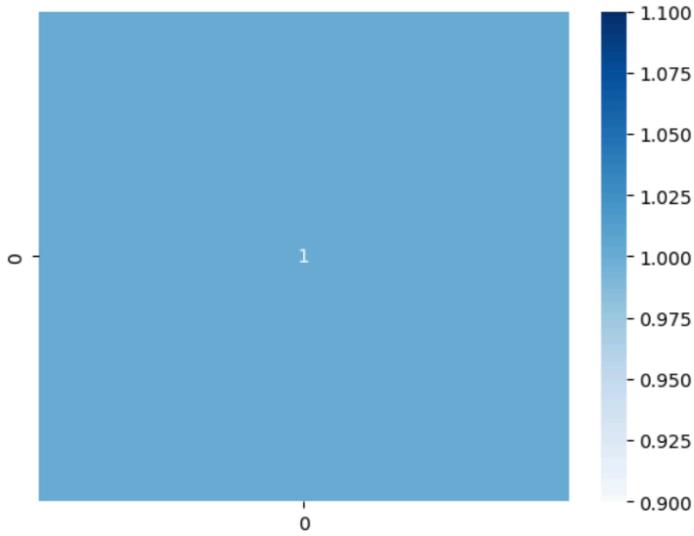

Fig 2: Confusion matrix of GaussianNB model

```
              precision    recall  f1-score   support

           0       0.00      0.00      0.00       0.0
           1       0.00      0.00      0.00       1.0

    accuracy                           0.00       1.0
   macro avg       0.00      0.00      0.00       1.0
weighted avg       0.00      0.00      0.00       1.0
```

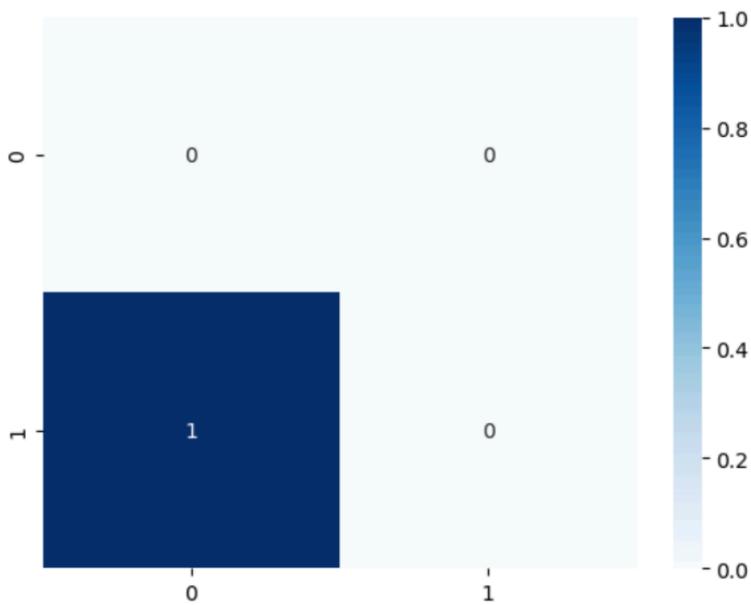

Fig 3: Confusion matrix of randomforest model

## 5.2 CONCLUSION

Sentiment analysis of thread reviews is an essential method for gaining insight into the opinions, preferences, and general levels of satisfaction of customers. Through the use of this technique, organizations may extract useful insights from unstructured text data, which can be applied to improve customer experiences, services, and products. Businesses may build a comprehensive and user-friendly sentiment analysis system by putting in place a solution that combines a Tableau Dashboard for data visualization with a Tkinter GUI for user interaction.

The Tkinter GUI offers a user-friendly interface that makes it simple for users to enter data, alter analytic settings, and communicate with the system. Because of the platform's intuitive design, users of all skill levels may easily traverse it, which promotes wider acceptance and utilization throughout departments within a company. Furthermore, the ability to interpret data in real time of the technology enables quick thread review analysis, guaranteeing timely insights that can support well-informed decision-making.

Conversely, Tableau Dashboard provides dynamic and interactive representations that make difficult sentiment data understandable and insightful. Users may efficiently study sentiment trends, find patterns, and spot anomalies thanks to the configurable dashboards, which aid in a deeper comprehension and interpretation of the data. Users may more quickly and effectively extract actionable insights thanks to this visual portrayal of sentiment analysis data, which improves understanding and engagement.